\documentclass{article}
\pdfoutput=1
\usepackage{lscape}
\usepackage{lipsum}
\usepackage{enumerate}
\usepackage{amsfonts}
\usepackage{amsmath}
\usepackage{comment}
\usepackage{textcomp}
\usepackage{amsthm, tikz}
\usepackage{amssymb}
\usepackage{color}
\usepackage{graphicx}
\usepackage{pdflscape}
\usepackage{afterpage}
\usepackage[matrix,arrow,curve]{xy}
\usepackage{array}
\usepackage{mathtools}

\usepackage{braids}

\makeatletter
\renewcommand*{\@cite@ofmt}{\bfseries\hbox}
\renewcommand{\L}{\mathcal{L}}
\newcommand{\K}{\mathcal{K}}
\newcommand{\R}{\mathcal{R}}
\newcommand{\M}{\mathcal{M}}
\makeatother

\def\be{\begin{eqnarray}}
\def\ee{\end{eqnarray}}

\hoffset= - 1.1in         

\begin{document}

\title{\vspace{0.1cm}{\Large {\bf New symmetries for the $U_q(sl_N)$ 6-j symbols \\ from \\ the Eigenvalue conjecture  }\vspace{.7cm}}
\author{
{\bf Andrey Morozov$^{a,b,c}$},
{\bf Alexey Sleptsov$^{a,b,c,d}$}}
\date{ }
}

\maketitle

\vspace{-5.5cm}

\begin{center}
\hfill IITP/TH-16/18\\
\hfill ITEP/TH-27/18\\
\end{center}

\vspace{3.9cm}

\begin{center}

$^a$ {\small {\it Institute for Information Transmission Problems, Moscow 127994, Russia}}\\
$^b$ {\small {\it Moscow Institute of Physics and Technology, Dolgoprudny 141701, Russia}}\\
$^c$ {\small {\it Institute for Theoretical and Experimental Physics, Moscow 117218, Russia}}\\
$^d$ {\small {\it Laboratory of Quantum Topology, Chelyabinsk State University, Chelyabinsk 454001, Russia }}
\end{center}

\vspace{1cm}

\begin{abstract}
In the present paper we discuss the eigenvalue conjecture, suggested in 2012, in the particular case of $U_q(sl_2)$. The eigenvalue conjecture provides a certain symmetry for Racah coefficients and we prove that \textbf{the eigenvalue conjecture is provided by the Regge symmetry} for $U_q(sl_2)$, when three representations coincide. This in perspective provides us a kind of generalization of the Regge symmetry to arbitrary $U_q(sl_N)$.

\end{abstract}


\vspace{.5cm}



\section{Introduction}
\setcounter{footnote}{1}

Racah coefficients (6-j symbols) are an important quantity in the theoretical and mathematical physics. They appear everywhere from quantum mechanics to the knot theory and integrable systems. Despite this there are a lot of open problems with Racah coefficients even for $U_q(sl_N)$. In the case of $U_q(sl_2)$ there is a general formula \cite{KirResh} for Racah coefficients. According to that formula 6-j symbols are described by the q-hypergeometric function $_4\Phi_3$. Using this fact the whole set of symmetries of $U_q(sl_2)$ Racah coefficients was described.

The next step was done in \cite{NRZ1}-\cite{MMSracah} by extending the explicit Racah formula to the symmetric representations of arbitrary $U_q(sl_N)$. It covers only so-called {\it exclusive} 6-j symbols, which appear in the arborecent links \cite{arbor1,arbor2} calculus coming from the Wess-Zumino-Novikov-Witten Confromal Field Theory consideration \cite{arborcalc1}-\cite{RTmod4}. Also exclusive 6-j symbols for the first non-symmetric representation $[2,1]$ was calculated in \cite{GuJ}.

The modern version of the Reshetikhin-Turaev approach provide a general method to compute colored HOMFLY polynomials for arbitrary link \cite{RTmod}.
Racah coefficients play the key role in this formalism, because they relate different $\mathcal{R}$-matrices. Some of the Racah coefficients appear in the Yang-Baxter equation together with $\mathcal{R}$-matrices. This led to the eigenvalue conjecture which relates the $\mathcal{R}$-matrix eigenvalues with the Racah coefficients. Eigenvalue conjecture, suggested in \cite{IMMMec},  is a very important concept which has a very useful practical applications. This conjecture states that Racah coefficients are fully defined by the set of normalised eigenvalues of the corresponding $\mathcal{R}$-matrices. In its strong form the exact form for Racah coefficients from the $\mathcal{R}$-matrix eigenvalues is provided, and it is known for the matrices of the size less or equal to $5\times 5$ \cite{IMMMec,Wenzl}. The explicit form for the Racah matrix of the size $6\times 6$ was proposed in \cite{MkrMM}. In its weak form it allows for example to calculate knot polynomials in any symmetric representation, expressing all the needed Racah coefficients through the known $U_q(sl_2)$ Racah coefficients \cite{DMMMRSS,LL}. Eigenvalue conjecture provides a certain symmetry for the Racah coefficients.

The eigenvalue conjecture gives rise to the question what happens with this symmetry in the well-known $U_q(sl_2)$ case. In the present paper we prove that:

\textbf{for $U_q(sl_2)$ Racah coefficients the eigenvalue conjecture is provided by the Regge symmetry.}

However, unlike the Regge symmetry, the eigenvalue conjecture is also formulated for arbitrary rank $U_q(sl_N)$. Thus one can hope that studies of the eigenvalue conjecture can give us some insights into the generalization of known symmetries from $U_q(sl_2)$ to $U_q(sl_N)$ case.

\bigskip

Our paper is organized as follows. In section 2 we give some basic definitions, briefly review the Reshetikhin-Turaev approach to quantum link invariants and formulate the eigenvalue conjecture as the symmetry property on the Racah matrices. In section 3 we remind the symmetry properties of 6-j symbols. In section 4 we prove that the eigenvalue conjecture is true in the case of $U_q(sl_2)$ if the three incoming representations are same. This particular case of 6-j symbols is enough to calculate colored HOMFLY polynomials of 3-strand knots and 3-strand links, whose components colored by the same representations. We also show that the Regge symmetry provide the eigenvalue conjecture in this case.  In section 5 we give some evidences that the eigenvalue conjecture correctly predicts the symmetry properties of the Racah matrices.

\section{Link invariants from quantum groups}
\label{s.int}
In this section we give a definiton of the colored HOMFLY polynomials of arbitrary links via Reshetikhin-Turaev approach based on the theory of quantum groups and quantum $\mathcal{R}$-matrix.

\subsection{$\mathcal{R}$-matrix and link invariants}
First of all, let us define quantum $\mathcal{R}$-matrices, which are associated with multicolored braid. Suppose that we have a braid with $m$ strands. We associate a finite-dimensional representation $R_i$ of the quantized universal enveloping algebra $U_q(sl_N)$ with $i$-th strand. Since we assume that $q$ is a nonzero complex number, which is not a root of unity, all finite-dimensional representations are representations of highest weights and they can be enumerated by Young diagrams. Due to this fact for simplicity we identify Young diagram and representation and use the same notations for both when it can be done without ambiguity.

\bigskip

$\bullet$ There exists universal R-matrix
\begin{equation}
\check{\mathcal{R}} = q^{\sum\limits_{i,j}C_{ij}^{-1}H_i\otimes H_j}
\prod_{\textrm{positive root }\beta} \exp_q
[( 1-q^{-1}) E_\beta\otimes F_\beta]\,.
\end{equation}
here $( C_{ij})$ is the Cartan matrix and $\{H_i,E_i,F_i\}$ are generators of $U_q(sl_N)$.

\bigskip

$\bullet$  If we define invertible linear operators by
\begin{equation}
\label{Rmat}
\mathcal{R}_i = 1_{V_1}\otimes1_{V_2}\otimes\ldots\otimes P \check{\mathcal{R}}_{i,i+1} \otimes\ldots\otimes1_{V_m} \ \in \text{End}(V_1\otimes\ldots,\otimes V_m),
\end{equation}
where $P(x\otimes y) = y\otimes x$ and $\check{\mathcal{R}}$ acts on two $U_q(sl_N)$-modules $V_i$ and $V_{i+1}$, then it is well known \cite{KirResh}, \cite{RJ} that $\mathcal{R}_1,\ldots,\mathcal{R}_{m-1}$ define a representation of the Artin's braid group $B_m$ on $m$ strands:
\begin{equation}
\begin{array}{rcl}
\pi: B_m  &\rightarrow&  \text{End}(V_1\otimes\ldots,\otimes V_m) \\
\pi(\sigma_i) &=& \mathcal{R}_i,
\end{array}
\end{equation}
where $\sigma_1,\ldots,\sigma_{m-1}$ are generators of the braid group $B_m$.
Graphically we can represent $\mathcal{R}_i$ as follows:

\begin{picture}(850,125)(-250,-93)

\put(-70,20){\line(1,0){90}}
\put(-35,0){\mbox{$\ldots$}}
\put(-70,-20){\line(1,0){35}}
\put(-70,-40){\line(1,0){35}}
\put(-15,-20){\line(1,0){35}}
\put(-15,-40){\line(1,0){35}}
\put(-35,-60){\mbox{$\ldots$}}
\put(-70,-80){\line(1,0){90}}
\put(-35,-20){\line(1,-1){20}}
\multiput(-15,-20)(-10,-10){2}{\line(-1,-1){10}}
\put(-130,-33){\mbox{$\mathcal{R}_i$ \ \ = \ \ }}
\put(-90,17){\mbox{$V_1$}}
\put(-90,-23){\mbox{$V_i$}}
\put(-90,-43){\mbox{$V_{i{+}1}$}}
\put(-90,-83){\mbox{$V_m$}}

\end{picture}

Clearly, inverse crossing is given by  $\mathcal{R}^{{-}1}_i$. Operators $\mathcal{R}_i$ satisfy  relations of the braid group $B_m$:
\begin{equation}
\begin{array}{lll}
\textbf{far commutativity} & \mathcal{R}_i\mathcal{R}_j = \mathcal{R}_j\mathcal{R}_i, & \text{for} \ |i-j|\neq 1 \\
\textbf{braiding relation} & \mathcal{R}_i\mathcal{R}_{i+1}\mathcal{R}_{i} = \mathcal{R}_{i+1}\mathcal{R}_i\mathcal{R}_{i+1}, & \text{for} \ i=1,\ldots,m-2
\end{array}
\end{equation}

Graphically, the braiding relation is nothing but the third Reidemeister move, while algebraically, it is a well-known \textit{quantum Yang-Baxter equation} on quantum $\mathcal{R}$-matrix.

\bigskip

$\bullet$ Alexander's theorem states that any link $\L$ in $\mathbb{R}^3$ can be obtained by a closure of the corresponding braid. Let $\L$ be an oriented link with $L$ components $\K_1,\ldots,\K_L$ colored by representations $R_1,\ldots,R_L$ and  $\beta_{\L} \in B_m$ is a some braid, which closure gives $\L$. Then according to Reshetikhin-Turaev approach \cite{RT0}-\cite{LiuPeng} the quantum group invariant, also known as colored HOMFLY polynomial, of the link $\L$ is defined as follows\footnote{The usual framing factor \cite{LiuPeng} in front of the quantum trace, which provide the invariance under the first Reidemeister move, we incorporate in the $\R$-matrix by modifying its eigenvalues (\ref{evR}).}:
\begin{equation}
\label{HMF1}
H_{R_1,\dots,R_L}^{\L} =  {}
_q\text{tr}_{V_1\otimes\dots\otimes V_m}\left( \, \pi(\beta_{\L}) \, \right),
\end{equation}
where $_q\text{tr}$ is a quantum trace.

\bigskip

$\bullet$ Although a  quantum trace is a standard notion in the theory of quantum groups \cite{Klimyk}, we give here some details, which are useful in Reshetikhin-Turaev approach.

Let $\rho$ denotes a half-sum of all positive roots of $sl_N$. There exists an element $K_{2\rho} \in U_q(sl_N)$ defined by
\begin{equation}
\begin{array}{l}
K_{2\rho} = K_1^{n_1} \, K_2^{n_2} \ldots K_{N-1}^{n_{N-1}}, \\
2\rho = \sum\limits_{i=1}^{N-1} n_i\alpha_i, \ n_i \in \mathbb{N}_0, \\
K_i = q^{H_i} \ \forall i=1,...,N-1,
\end{array}
\end{equation}
where $\alpha_i$ are simple roots. Then for every $z \in \text{End}(V)$ one has the quantum trace
\begin{equation}
\label{qtr}
_q\text{tr}_V(z) = \text{tr}_V(zK_{2\rho}).
\end{equation}

\bigskip

$\bullet$ Let us expand $V_1\otimes V_2\otimes \ldots \otimes V_m$ into a direct sum of irreducible representations:
\begin{equation}
\label{irrdec}
\bigotimes_i V_i = \bigoplus_{\mu} \mathcal{M}_\mu \otimes Q_{\mu},
\end{equation}
where $Q_{\mu}$ is an irreducible representation and $\mathcal{M}_\mu$ is the subspace of highest weight vectors with highest weights\footnote{
	Recall, that if $\mu$ is a Young diagram $\mu=\{\mu_1\geq\mu_2\geq\ldots,\mu_l>0\}$, then the highest weights $\vec{\omega}$ of the corrsponding representation are $\omega_i= \mu_i-\mu_{i+1} \ \forall \, i=1,\dots,l$, and vice versa $\mu_i = \sum_{k=i}^l \, \omega_k$.}
corresponding to a Young diagram $\mu$. The dimension of the space $\mathcal{M}_\mu$ is called \textit{the multiplicity} of representation $Q_{\mu}$. 

Let us evaluate quantum trace (\ref{qtr}) on such decomposition (\ref{irrdec}). Since $\mathcal{R}$-matrix commutes with any element from $U_q(sl_N)$, then $\mathcal{R}$-matrix gets a block structure corresponding to the decomposition on irreducible components (\ref{irrdec}), i.e. it does not mix vectors from different representations. Futhermore, $\mathcal{R}$-matrix acts on $Q_{\mu}$ as an identity operator, while on $\mathcal{M}_\mu$ it acts nontrivially. The element $K_{2\rho}$ acts diagonally on $Q_{\mu}$, while on $\mathcal{M}_\mu$ it acts identically, because there are highest weight vectors with the same weight $\vec{\omega}(\mu)$. Therefore, the decomposition (\ref{irrdec}) implies for the colored HOMFLY (\ref{HMF1}) the following
\begin{equation}
\begin{array}{ccc}
H_{R_1,\dots,R_L}^{\L} \left(q,\,A=q^N \right)
\ = \
\text{tr}_{V_1\otimes\dots\otimes V_m}\left( \, \pi(\beta_{\L}) \, K_{2\rho} \, \right) \ = \  \sum \limits_{\mu\, \vdash \sum |R_i|} \text{tr}_{\mathcal{M}_\mu}\left( \, \pi(\beta_{\L}) \, \right) \cdot \text{tr}_{Q_{\mu}}\left( \, K_{2\rho} \, \right) = \\ \\
=  \sum\limits_{\mu\, \vdash \sum |R_i|} \text{tr}_{\mathcal{M}_\mu}\left( \, \pi(\beta_{\L}) \, \right) \cdot \text{qdim}_{\mu},
\label{linv}
\end{array}
\end{equation}
where $\text{qdim}_{\mu}$ is a quantum dimension of the representation $Q_{\mu}$ explicitly given in terms of Schur polynomials \cite{Lin}:
\begin{equation}
\text{qdim}_{\mu} = s_{\mu}\left( x_1,\ldots,x_N \right){\Big|_{x_i=q^{N+1-2i}}} = s_{\mu}^{}(p_1,\dots,p_N)\Big|_{p_k=p_k^{*}} \equiv s_{\mu}^{*}(A,q),
\end{equation}
where $p_k=\sum\limits_{i=1}^{N} x_i^k$ and $p_k^* = \frac{A^k-A^{-k}}{q^k-q^{-k}}$.

\subsection{Quantum Racah coefficients  (6-j symbols)}
In this subsection we define quantum Racah coefficients, also known as 6-j symbols. 

Consider three finite-dimensional irreducible representations $R_{i_1},\ R_{i_2}, \ R_{i_3}$ of highest weights of $U_q(sl_N)$. Since the tensor product of these representations is associative, one has a natural isomorphism:
\begin{equation}
\label{asis}
\left( R_{i_1}\otimes R_{i_2} \right)\otimes R_{i_3} \rightarrow R_{i_1} \otimes \left( R_{i_2}\otimes R_{i_3} \right).
\end{equation}
Expand tensor product of two representations into irreducible components like in (\ref{irrdec}):
\begin{equation}
\begin{array}{l}
R_{i_1}\otimes R_{i_2} = \bigoplus_{\mu} \mathcal{M}^{i_1,i_2}_\mu \otimes X_{\mu}, \\
R_{i_2}\otimes R_{i_3} = \bigoplus_{\nu} \mathcal{M}^{i_2,i_3}_\nu \otimes Y_{\nu},
\end{array}
\end{equation}
and once again
\begin{equation}
\begin{array}{l}
\left( R_{i_1}\otimes R_{i_2} \right)\otimes R_{i_3} = \bigoplus_{\mu,\xi} \mathcal{M}^{i_1,i_2}_\mu \otimes \mathcal{M}^{\mu,i_3}_\xi \otimes R_{\xi}, \\
R_{i_1} \otimes \left( R_{i_2}\otimes R_{i_3} \right) = \bigoplus_{\nu,\xi} \mathcal{M}^{i_1,\nu}_\xi \otimes \mathcal{M}^{i_2,i_3}_\nu \otimes R_{\xi}.
\end{array}
\end{equation}
Then the associativity isomorphism (\ref{asis}) implies
\begin{equation}
\Phi^{i_1,i_2}_{i_3,\xi}: \ \bigoplus_{\mu} \mathcal{M}^{i_1,i_2}_\mu \otimes \mathcal{M}^{\mu,i_3}_\xi \ \rightarrow \ \bigoplus_{\nu} \mathcal{M}^{i_1,\nu}_\xi \otimes \mathcal{M}^{i_2,i_3}_\nu,
\end{equation}
and the Racah coefficients are defined as the  map components:
\begin{equation}
U\left[\begin{array}{cc|c} i_1&i_2& \mu \\ i_3&\xi &\nu\end{array}\right] = \left( \Phi^{i_1,i_2}_{i_3,\xi} \right)_{\mu,\nu}.
\end{equation}
Since this paper is devoted to Racah coefficients, then let us change some notations to make them more convenient and obvious. Instead of indices $i_1, \ i_2, i_3$ and $\xi$ we shall use representations $R_1, \ R_2, \ R_3$ and $R_4$, and instead of intermediate indices $\mu$ and $\nu$ we use $X \in R_1\otimes R_2$ and $Y \in R_2\otimes R_3$. Thus, we denote Racah matrix by
\begin{equation}
U\left[\begin{array}{cc} R_1&R_2 \\ R_3&R_4 \end{array}\right]
\end{equation}
and Racah coefficient by
\begin{equation}
U\left[\begin{array}{cc|c} R_1&R_2& X \\ R_3&R_4 &Y\end{array}\right].
\end{equation}
Usually quantum 6-j symbols have different normalisation than Racah coefficients, but for our purposes such normalisation is not important, hence we assume that they coincide:
\be
\left\{\begin{array}{ccc}  R_1 & R_2 & X \\ R_3 & R_4 & Y \end{array}\right\} \equiv
U\left[\begin{array}{cc|c} R_1&R_2& X \\ R_3&R_4 &Y\end{array}\right]
\ee

Graphically Racah matrix can be represented as follows

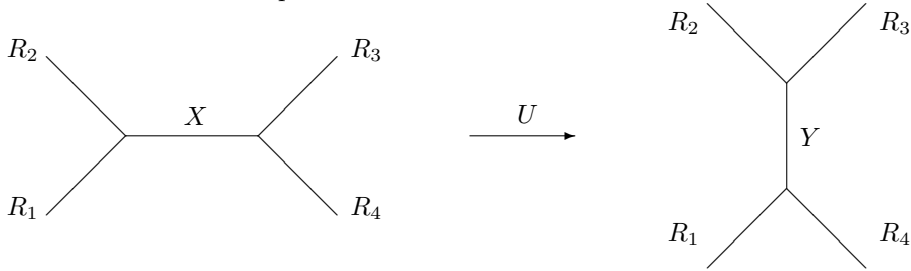
\begin{picture}(240,110)(-100,-60)
\put(0,0){\line(1,0){50}}
\put(0,0){\line(-1,1){30}}
\put(0,0){\line(-1,-1){30}}
\put(50,0){\line(1,1){30}}
\put(50,0){\line(1,-1){30}}
\put(-45,-30){\mbox{$R_1$}}
\put(-45,30){\mbox{$R_2$}}
\put(85,-30){\mbox{$R_4$}}
\put(85,30){\mbox{$R_3$}}
\put(22,4){\mbox{$X$}}
\put(130,0){\vector(1,0){40}}
\put(148,5){\mbox{$U$}}
\put(250,0){
	\put(0,-20){\line(0,1){40}}
	\put(0,-20){\line(-1,-1){30}}
	\put(0,-20){\line(1,-1){30}}
	\put(0,20){\line(1,1){30}}
	\put(0,20){\line(-1,1){30}}
	\put(-45,-40){\mbox{$R_1$}}
	\put(-45,40){\mbox{$R_2$}}
	\put(35,40){\mbox{$R_3$}}
	\put(35,-40){\mbox{$R_4$}}
	\put(5,-4){\mbox{$Y$}}
}
\end{picture}

Their explicit calculation in great details through highest weight vectors for different representation can be found in \cite{MMMS21}-\cite{China}.

\subsection{$\mathcal{R}$-matrices via Racah matrices}
Now we describe how quantum Racah matrices relate different $\mathcal{R}$-matrices (\ref{Rmat}) with each other \cite{KirResh}.

$\bullet$ Let us choose the basis on $R_1\otimes\dots\otimes R_m$, which corresponds to
the following order in the tensor product:
\begin{equation}
B_{12,3,..,m} := \left(\ldots\Big(\,\left(  R_1\otimes R_2 \right)\otimes R_3 \Big) \otimes \ldots \right) \otimes R_m.
\end{equation}
Then the matrix $\R_1$ in this basis gets a block form on each space $\M_{\mu}$ from (\ref{irrdec}). Different blocks corresponds to different representations $X_{\alpha}$ in the decomposition
\begin{equation}
R_1 \otimes R_2 = \bigoplus_{\mu} \mathcal{M}^{1,2}_\alpha \otimes X_{\alpha}
\end{equation}
as we discussed after formula (\ref{irrdec}). The size of the block, which correspond to $X_{\alpha}$, is dim$\,\mathcal{M}^{1,2}_\alpha \times$ dim$\,\mathcal{M}^{1,2}_\alpha$. Therefore, if dim$\,\mathcal{M}^{1,2}_\alpha > 1$, then we need to rotate the basis additionally (those components, which correspond to $\alpha$) to diagonalize the matrix $\R_1$, but it is always possible to do. Thus, in the basis $B_{12,3,..,m}$ it is always possible to diagonalize the matrix $\R_1$.

\bigskip

$\bullet$ In order to diagonalize the matrix $\mathcal{R}_2$ we have to repeat the same procedure but for the basis corresponding to
\begin{equation}
B_{1,23,..,m} := \left(\ldots\Big(  R_1\otimes \left(R_2 \otimes R_3 \right) \Big)\otimes \ldots \right)\otimes R_m.
\end{equation}
Therefore, in order to diagonalize the matrix $\mathcal{R}_2$ one should make the basis transformation with the help of Racah matrix:
\begin{equation}
\label{R2}
\mathcal{R}_2 = U^{\dagger}\left[\begin{array}{cc} R_1&R_3 \\ R_2&R_4 \end{array}\right] \cdot \text{diag}\left( \lambda_{\mathcal{R}_2} \right) \cdot U\left[\begin{array}{cc} R_1&R_2 \\ R_3&R_4 \end{array}\right],
\end{equation}
where $\lambda_{\mathcal{R}_2}$ mean eigenvalues of the matrix $\mathcal{R}_2$ and $U$ are corresponding Racah matrices.

Similarly we can diagonalize all $\mathcal{R}$-matrices. So, in order to evaluate link invariant (\ref{linv}) we need to find eigenvalues of $\mathcal{R}$-matrices and Racah coefficients. While eigenvalues of $\mathcal{R}$-matrices are known explicitly and we give answer for arbitrary representations below, Racah coefficients are known only for a few simple cases. Their calculation is puzzling and hard problem standing over 50 years both for quantum and classical groups, solved only for $U_q(sl_2)$ case.


\subsection{Eigenvalues of $\R$-matrix}

Now let us consider tensor product of two representations  $R$ and  $R$ and irreducible representations $X_{\alpha}$, which occur in their decomposition into irreducible components:
\begin{equation}
R\otimes R = \bigoplus_{\alpha} X_{\alpha}.
\end{equation}
In this sum we allow repeated summands. Then according to \cite{GZ, Klimyk} eigenvalues of universal R-matrix are
\begin{equation}
\label{evR}
\begin{array}{l}
\lambda_{X_{\alpha}}=\epsilon_{X_{\alpha}} q^{\varkappa(X_{\alpha})-4\varkappa({R})-|R|N}
\end{array}
\end{equation}

where $\varkappa(X_{\alpha})$ defined by $\sum_{(i,j)\in \alpha} \, (i-j)$, and  $\epsilon_{X_{\alpha}}=\pm 1$ is a sign, which depends on whether highest weight vectors of the representations are \textit{symmetric} or \textit{antisymmetric} under permutation of two representations $R$ and $R$. These two types of representations $X_{\alpha}$ are said to belong to either symmetric or antisymmetric squares of the representation $R$.

\

Finally we define {\it normalised} diagonal $\mathcal{R}$-matrix by the following conditions
\begin{enumerate}
\item $\prod_i {\lambda}_i = 1$,
\item $\log_q |\lambda_1| \geq \log_q |\lambda_2| \geq \ldots$,
\item if $\log_q |\lambda_i| = \log_q |\lambda_{i+1}|$, then we put eigenvalues with positive signs first,
\item $\text{sgn}(\lambda_1) > 0$.
\end{enumerate}

\subsection{Eigenvalue conjecture}
\label{evcj}
In \cite{IMMMec} the eigenvalue conjecture was suggested. It relates quantum Racah coefficients with eigenvalues of $\mathcal{R}$-matrices. Since in this paper we consider 3-strand braids only, then we give a formulation for this case only. In the simplest case of $3$-strand knots the eigenvalue conjecture states that the Racah matrices are fully-defined by the set of normalised eigenvalues of the corresponding $\mathcal{R}$-matrix. If we consider $3$-strand braid with representations $R_1=R_2=R_3=R$. Then there is the unique diagonal $\mathcal{R}$-matrices, say, $\mathcal{R}_{1}$ corresponding to basis $B_{12,3}$.

The source of the eigenvalue conjecture is the Yang-Baxter equation, which for the 3-strand braid reads as
\be
\R_1\R_2\R_1=\R_2\R_1\R_2
\ee
Substituting formula \eqref{R2} in this equation we get
\be
\label{req}
\R_1\cdot U^{\dagger} \R_1  U \cdot \R_1  =  U^{\dagger} \R_1  U \cdot \R_1 \cdot U^{\dagger} \R_1  U.
\ee
The latter equation is invariant under a scalar multiplication of the $\R_1$-matrix and permutations of diagonal elements. Therefore, if we have $\R_1$ and $\tilde{\R}_1$, which coincide after the normalisation procedure, then Racah matrices $U$ and $\tilde U$ satisfy same equation \eqref{req}. A priori there is no reason this equation (in more exact terms it is a system on nonlinear equations) has always a unique solution corresponding to the quantum group. Nevertheless we formulate the following conjecture.

\textbf{Conjecture:} \textit{Let we have two sets $\{R,R,R,Q;\, \R_1, U\}$ and $\{\tilde R,\tilde R,\tilde R,\tilde Q;\,\tilde  \R_1,\tilde  U\}$, where $R,\tilde R,Q,\tilde Q$ are representations as follows $R^{\otimes 3}\rightarrow Q$, $\tilde R^{\otimes 3}\rightarrow \tilde Q$, matrices $\R_1$ and $\tilde R_1$ are the corresponding diagonal $\R$-matrices,  $U$ and $\tilde U$ are the corresponding Racah matrices. Let normalised $R_1$ concides with normalised $\tilde \R_1$. Then $U=\tilde U$.   }

\textbf{Remark:} If the normalised $R_1$ concides with the normalised $\tilde \R_1^{-1}$, then we conjecture $U=\tilde U^{-1}$.

\section{The symmetry properties of 6-j symbols}

In this section we briefly review the symmetry properties of quantum 6-j symbols.

{$\bullet$ $\rm \bf U_q(sl_2)$} In this case all symmetries are known due to representation of 6-j symbols in terms of q-hypergeometric function. The symmetry group contains 144 elements, the full tetrahedral group $S_4$ is its subgroup. It describes \textit{the tetrahedral} symmetry, which can be represented in a pictorial way by the associating a corresponding tetrahedron with the 6-j symbol:

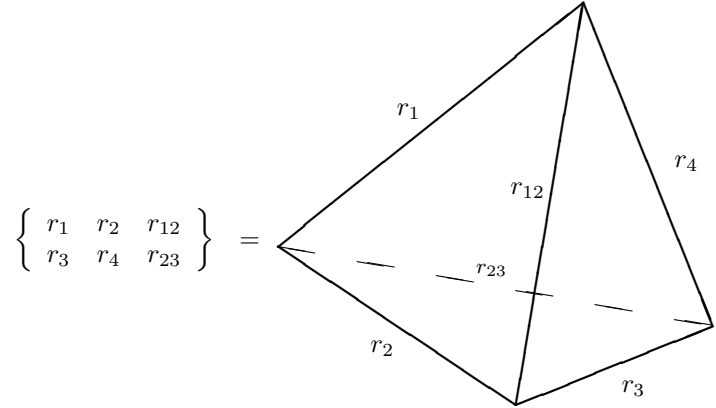
\begin{picture}(850,160)(-250,-25)
\put(-200,40){\mbox{$\left\{\begin{array}{ccc}
		r_1 & r_2 & r_{12}
		\\
		r_3 & r_4 & r_{23}
		\end{array}\right\} \ \ = $}}
\multiput(-100,40)(30,-5.5){6}{\line(6,-1){14}}
\thicklines
\put(-100,40){\line(3,-2){90}}
\put(-10,-20){\line(1,6){25.4}}
\put(-100,40){\line(5,4){115.5}}
\put(-10,-20){\line(5,2){74.5}}
\put(15.5,132.5){\line(2,-5){49}}
\put(-55,90){\mbox{$r_1$}}
\put(-65,0){\mbox{$r_2$}}
\put(-12,60){\mbox{$r_{12}$}}
\put(30,-15){\mbox{$r_3$}}
\put(50,70){\mbox{$r_4$}}
\footnotesize{
\put(-25,30){\mbox{$\tiny r_{23}$}}}
\end{picture}

Other symmetries are given by the q-analog of \textit{the Regge} symmetry. Let $p=\frac{1}{2}\left( r_1+r_2+r_3+r_4 \right)$, then the Regge symmetry holds:
\be
\label{Regge}
\left\{\begin{array}{ccc}
	r_1 & r_2 & r_{12}
	\\
	r_3 & r_4 & r_{23}
\end{array}\right\}  =
\left\{\begin{array}{ccc}
p-r_1 & p-r_2 & r_{12}
\\
p-r_3 & p-r_4 & r_{23}
\end{array}\right\}
\ee

\bigskip

Alternatively, we can represent all symmetries in the following way. Let us take the 6-j symbol
\begin{equation}
\left\{\begin{array}{ccc}
r_1 & r_2 & r_{12}
\\
r_3 & r_4 & r_{23}
\end{array}\right\}
\end{equation}
and reexpress it through a different set of parameters:
\begin{equation}
\left[\begin{array}{ccc}
A & B & C
\\
\alpha & \beta & \gamma
\end{array}\right]
:=
\left\{\begin{array}{ccc}
\frac{1}{2}(A+\alpha) & \frac{1}{2}(B+\beta) & \frac{1}{2}(C+\gamma) \phantom{\dfrac{1}{8}}
\\
\frac{1}{2}(A-\alpha) & \frac{1}{2}(B-\beta) & \frac{1}{2}(C-\gamma)
\end{array} \hspace{-2mm} \right\}
\end{equation}
Then the 6-j symbol is  invariant under separate permutations of $\{A,B,C\}$ alone, separate permutations of  $\{\alpha,\beta,\gamma\}$ alone and separate change in sign of any pair of  $\{\alpha,\beta,\gamma\}$.

\bigskip

{$\bullet$ $\rm \bf U_q(sl_N)$} In this case we know only about the tetrahedral symmetry \cite{GuJ,Lienert,Pan}. The counterpart of the Regge symmetry is unknown at the present moment.

\section{Eigenvalue conjecture for $U_q(sl_2)$}

In this section we prove that the eigenvalue conjecture is true if $R_1=R_2=R_3=[r]$ for $U_q(sl_2)$. Moreover, we show that it follows from the Regge symmetry.

Since $R_4 \in R_1\otimes R_2\otimes R_3$ the corresponding Young diagram according to the Littlewood-Richardson rule can take the form $[2m], \ m=0\ldots \frac{3}{2}r$. We consider two cases $R_4=[r+2k], \ k=0\ldots r$ and $R_4=[r-2k], \ k=0\ldots \lfloor\frac{r}{2}\rfloor$ separately.

\

\noindent
\textbf{1.} $R_4=[r+2k]$, where $k=0\ldots r$.

To understand which eigenvalues correspond to the Racah matrix we should find all representations $X \in [r]\otimes [r]$, which contributes in this case. In general these representations can be denoted as $[2r-2k_1]$. Then there are the following restrictions:
\begin{equation}
\begin{array}{lcl}
\ [2r-2k_1]\in [r]\otimes [r]           & & 0\leq k_1 \leq r
\\
& \Rightarrow                             &
\\
\ [r+2k]\in [2r-2k_1]\otimes [r]  & & k_1\leq r-k
\end{array}
\end{equation}
Thus it comes to $0\leq k_1\leq r-k$. Therefore, the corresponding 6-j symbols are
\be
\label{rac1}
\left\{\begin{array}{ccc}
r & r           & 2r-2k_1
\\
r & r+2k  & 2r-2k_2
\end{array}\right\},\ \ \ k_1,k_2=0\ldots r-k.
\ee
The size of the Racah matrix is given by $d=r-k+1$. Since the eigenvalue conjecture implies to fix the size of the Racah matrices, then we must vary representation $r$ in \eqref{rac1} in order to consider nontrivial cases.  However the eigenvalue conjecture does not give us something nontrivial in this case, because there are no two different representations $r$ and $R$ such that the corresponding normalised eigenvalues coincide. It follows from a staraightforward calculation of a difference of the powers between two adjacent eigenvalues:
\be
\varkappa([2r-2k_1]) - \varkappa([2r-2k_1-2]) = 4r-4k_1-3, \qquad \forall k_1=0\ldots r-k-2.
\ee

\

\noindent
\textbf{2.} $R_4=[r-2k]$, where $k=0\ldots \lfloor\frac{r}{2}\rfloor$.

Again let us look at the representations from the tensor square:
\begin{equation}
\begin{array}{lcl}
\ [2r-2k_1]\in r\otimes r           & & 0\leq k_1 \leq r
\\
& \Rightarrow                               &
\\
\ [r-2k]\in [2r-2k_1]\otimes r  & & |r-2k_1| \leq r-2k \leq |3r-2k_1|
\end{array}
\end{equation}
This gives $k\leq k_1\leq r-k$. Therefore, the corresponding Racah coefficients are
\begin{equation}
\left\{\begin{array}{ccc}
r & r           & 2r-2k_1
\\
r & r-2k  & 2r-2k_2
\end{array}\right\},\ \ \ k_1,k_2=k\ldots r-k
\end{equation}
The size of the Racah matrix is given by $d=r-2k+1$. Similar reasoning, which we use above, shows that the eigenvalue conjecture says nothing about type 2 representations.

\

\noindent
\textbf{3.} Let us now consider a type 1 Racah matrix of the size $d\times d$  and a type 2 matrix of the same size and apply the eigenvalue conjecture. The corresponding sets of representations in the tensor square are
\begin{equation}
\begin{array}{c}
\ [2r],\ [2r-2], \, \ldots ,\ [2r-2d+2]
\\
\ [R+d-1], \ [R+d-3], \, \ldots,  \ [R-d+1]
\end{array}
\end{equation}
If $R+d-1=2r$ then eigenvalues coincide. So eigenvalue conjecture tells us that the following Racah coefficients should coincide. For a type 1 matrix and a representation $r$, substituting $k=r+1-d$, we have:
\begin{equation}
\label{rrac}
\boxed{
\left\{\begin{array}{ccc}
r & r           & 2r-2k_1
\\
r & 3r-2d+2  & 2r-2k_2
\end{array}\right\},\ \ \ k_1,k_2=0\ldots d-1
}
\end{equation}
For a type 2 matrix and a representation $R$, substituting $R=2r-d+1$ and $k=\frac{R-d+1}{2}=r-d+1$:
\be
\label{Rrac}
\begin{array}{l}
\left\{\begin{array}{ccc}
2r-d+1 & 2r-d+1           & 4r-2d+2-2k_1
\\
2r-d+1 & d-1  & 4r-2d+2-2k_2
\end{array}\right\},\ \ \ k_1,k_2=r-d+1\ldots r
=  \\ \\
=
\boxed{
\left\{\begin{array}{ccc}
2r-d+1 & 2r-d+1           & 2r-2k_1
\\
2r-d+1 & d-1  & 2r-2k_2
\end{array}\right\},\ \ \ k_1,k_2=0\ldots d-1
}
\end{array}
\ee

Thus, the eigenvalue conjecture predicts that 6-j symbol \eqref{rrac} coincides with \eqref{Rrac}. With the help of the Regge symmetry we can easily check that it is true. Indeed, the Regge symmetry \eqref{Regge} applied to the 6-j symbols \eqref{rrac} immediately gives 6-j symbols \eqref{Rrac}.

\section{Eigenvalue conjecture for $U_q(sl_N)$}
The eigenvalue conjecture also predicts the symmetry between different Racah matrices for the quantum group of the rank higher than 2. Such cases are more involved because they usually contain multiplicities. There is no analytical description of 6-j symbols like for $U_q(sl_2)$, consequently we can only perform some checks of the eigenvalue conjecture.

$\bullet$ In the papers \cite{IMMMec,Wenzl} there were found general nontrivial solutions of the Yang-Baxter equation \eqref{req} if the size of the R-matrix not greater than 5 and all its eigenvalues are different. These solutions provide explicit formulas for matrix elements (i.e. for 6-j symbols) in terms of algebraic functions depending on eigenvalues. Therefore, these formulas provide the proof of the eigenvalue conjecture for this particular class of matrices. Note these formulas are valid for any rank $N$ of the algebra $U_q(sl_N)$.

$\bullet$ If the size of the Racah matrix is greater than 5, then there are no explicit solutions of the Yang-Baxter equation. However we manage to present some particular examples of the Racah matrices, which confirm the eigenvalue conjecture. These examples we computed in the series of papers \cite{MMMS21}-\cite{China}:
\be
U\left[\begin{array}{cc} [3,3] & [3,3] \\ {[}3,3] & [7,5,3,3] \end{array}\right]  &=&
U\left[\begin{array}{cc} [3,3] & [3,3] \\ {[}3,3] & [6,6,4,2]  \end{array}\right], \qquad 6\times 6, \ \text{multiplicity free} \\
U\left[\begin{array}{cc} [3,3] & [3,3] \\ {[}3,3] & [8,5,3,2] \end{array}\right]  &=&
U\left[\begin{array}{cc} [3,3] & [3,3] \\ {[}3,3] & [7,6,4,1] \end{array}\right],  \qquad 6\times 6, \ \text{double multiplicity} \\
U\left[\begin{array}{cc} [4,2] & [4,2] \\ {[}4,2] & [5,5,3,3,2] \end{array}\right]  &=&
U\left[\begin{array}{cc} [4,2] & [4,2] \\ {[}4,2] & [6,4,3,3,1] \end{array}\right],  \quad 9\times 9, \ \text{two double multiplicities}.
\ee

Even these few examples give us a hope that further studies of the eigenvalue conjecture will provide some general symmetry of the Racah coefficients for any $U_q(sl_N)$ group. However at the moment some general expressions of the eigenvalue conjecture are lacking and this remains to be done.

Based on the previous section we can provide a $U_q(sl_N)$ version of the same relation. The formulae (\ref{rrac}) and (\ref{Rrac}) become:
\begin{equation}
\begin{array}{r}
\left[\begin{array}{ccc}
[2r-N+1] & [2r-N+1]           & [3r-N+1-k_1,k_1+r-N+1]
\\
{[}2r-N+1] & [3r-N+1,3r-2N+2]  & [3r-N+1-k_2,k_2+r-N+1]
\end{array}\right]
=
\\
=
\left[\begin{array}{ccc}
[r] & [r]           & [2r-k_1,k_1]
\\
{[}r] & [3r-N+1,N-1]  & [2r-k_2,k_2]
\end{array}\right] ,\ \ \ k_1,k_2=0\ldots N-1
\end{array}
\end{equation}
We hope that this formula is just the beginning of the formulation of Racah coefficients symmetries in the $U_q(sl_N)$ case.


\section*{Acknowledgements}
This work was funded by the Russian Science Foundation (Grant No.16-11-10291).

\end{document}